\documentclass[aps,prl,twocolumn,superscriptaddress,longbibliography]{revtex4-2}
\usepackage{graphicx}
\usepackage{latexsym}
\usepackage{amssymb}
\usepackage{amsmath}
\usepackage{amsfonts}
\usepackage{upgreek}
\usepackage{subfigure}
\usepackage{bm}
\usepackage{bbold}
\usepackage{verbatim}
\usepackage[unicode=true,
 bookmarks=false,
 breaklinks=false,pdfborder={0 0 1},backref=false,colorlinks=true]
 {hyperref}
\hypersetup{
 linkcolor=[rgb]{0,0,1},citecolor=[rgb]{0,0,1},urlcolor=[rgb]{0,0,1}}
\usepackage{multirow}
\usepackage{color}
\usepackage{comment}
\usepackage{xcolor}
\usepackage{soul}
\usepackage{tikz}

\usepackage{braket}

\begin{document}

%\begin{comment}

\title{Robust stripes in the mixed-dimensional $t-J$ model}

\author{Henning Schl\"omer}
\affiliation{Department of Physics and Arnold Sommerfeld Center for Theoretical Physics (ASC), Ludwig-Maximilians-Universit\"at M\"unchen, Theresienstr. 37, M\"unchen D-80333, Germany}
\affiliation{Munich Center for Quantum Science and Technology (MCQST), Schellingstr. 4, D-80799 M\"unchen, Germany}
\author{Annabelle Bohrdt}
\affiliation{Department of Physics, Harvard University, Cambridge, Massachusetts 02138, USA}
\affiliation{ITAMP, Harvard-Smithsonian Center for Astrophysics, Cambridge, MA, USA}
\author{Lode Pollet}
\affiliation{Department of Physics and Arnold Sommerfeld Center for Theoretical Physics (ASC), Ludwig-Maximilians-Universit\"at M\"unchen, Theresienstr. 37, M\"unchen D-80333, Germany}
\affiliation{Munich Center for Quantum Science and Technology (MCQST), Schellingstr. 4, D-80799 M\"unchen, Germany}
\author{Ulrich Schollw\"ock}
\affiliation{Department of Physics and Arnold Sommerfeld Center for Theoretical Physics (ASC), Ludwig-Maximilians-Universit\"at M\"unchen, Theresienstr. 37, M\"unchen D-80333, Germany}
\affiliation{Munich Center for Quantum Science and Technology (MCQST), Schellingstr. 4, D-80799 M\"unchen, Germany}
\author{Fabian Grusdt}
\affiliation{Department of Physics and Arnold Sommerfeld Center for Theoretical Physics (ASC), Ludwig-Maximilians-Universit\"at M\"unchen, Theresienstr. 37, M\"unchen D-80333, Germany}
\affiliation{Munich Center for Quantum Science and Technology (MCQST), Schellingstr. 4, D-80799 M\"unchen, Germany}

\date{\today}
\begin{abstract}
Microscopically understanding competing orders in strongly correlated systems is a key challenge in modern quantum many-body physics. For example, the origin of stripe order and its relation to pairing in the Fermi-Hubbard model remains one of the central questions, and may help to understand the origin of high-temperature superconductivity in cuprates. Here, we analyze stripe formation in the doped mixed-dimensional (mixD) variant of the $t-J$ model, where charge carriers are restricted to move only in one direction, whereas magnetic $\mathrm{SU}(2)$ interactions are two-dimensional. Using the density matrix renormalization group at finite temperature, we find a stable vertical stripe phase in the absence of pairing, featuring incommensurate magnetic order and long-range charge density wave profiles over a wide range of dopings. We find high critical temperatures on the order of the magnetic coupling $\sim J/2$, hence being within reach of current quantum simulators. Snapshots of the many-body state, accessible to quantum simulators, reveal hidden spin correlations in the mixD setting, whereby antiferromagnetic correlations are enhanced when considering purely the magnetic background. The proposed model can be viewed as realizing a parent Hamiltonian of the stripe phase, whose hidden spin correlations lead to the predicted resilience against quantum and thermal fluctuations.  
 
\end{abstract}
\maketitle

\textit{Introduction.---} The interplay of spin and motional degrees of freedom is at the heart of many strongly correlated quantum materials, leading to a plethora of interacting many-body phases. 
Microscopically understanding the competition between hole pairing and inhomogeneous stripe order in the paradigmatic Fermi-Hubbard (FH) model constitutes one of the central challenges in modern many-body physics, which may help to reveal the origin of high-temperature superconductivity~\cite{Darmawan2018, emery1999stripe, Himeda2002, BIANCONI20001719, TRANQUADA20121771}. In early experiments of cuprates and modern numerical studies of the FH model, the emergence of stripe order at low temperatures has been widely established~\cite{George_stripes, Kivelson_stripes, Machida_stripes, Zaanen_stripes, White_stripes, White_stripes2, Hager_stripes, Ehlers_stripes, Jiang_hubbard_pd, Huang2022}. However, it remains an open question whether pairing competes with stripes or if the two effects are different manifestations with a common origin~\cite{Corboz_comp, Zheng_comp, Qin2022, Qin_absence_SC}. 

Analog quantum simulation, e.g. via ultracold atoms, can help unveil the microscopic mechanisms underlying such strongly correlated phases~\cite{Bloch2012, Bloch2008, Gross2017, Schafer2020, Koepsell_science, Zohar2015, Cheuk2015, Omran2015, Parsons2015, Mazurenko_AFM}. In particular, recent advances allow for an implementation and experimental exploration of the FH model in ultracold atom setups~\cite{Hart2015, Tarruell, Chiu2019, Salfi2016,Esslinger2010,Hilker2017, Cocchi2016, Hirthe2022, Bohrdt2020}. However, studying stripe order with quantum simulators remains an open challenge, partly due to low critical temperatures~\cite{Wietek_stripes} and close degeneracies of different stripe fillings and paired states~\cite{Zheng2017}.

Here, we propose a parent Hamiltonian whose ground state forms a robust stripe phase over a wide range of dopings. The model assumes mixed dimensionality (mixD), whereby motional degrees of freedom are restricted to be one-dimensional (1D), but spin-superexchange interactions are two-dimensional (2D). As a result, a similar pairing mechanism as proposed recently in mixD bilayer antiferromagnets (AFMs)~\cite{bohrdt2021strong} allows holes to form stable vertical stripes. 

In this letter, we focus on general physical aspects of stripe order in the mixD setting. In particular, we analyze the order for various hole densities and map out the phase diagram. We predict high critical temperatures for stripe formation, rendering the stripe phase readily observable in ultracold atom experiments with optical lattices.

Our results shed new light on the long-standing question about the interplay of pairing and stripe formation. We find that the origin of stripes are hidden AFM correlations, making the stripe phase remarkably robust against thermal and quantum fluctuations. At the same time, intra-leg pairing of two holes -- the analog of a superconducting state -- is strongly suppressed. The model is closely related to the celebrated 2D $t-J$ model, offering an adiabatic route to gain new insights into stripes and their related phases in the FH model.

We use finite temperature density matrix renormalization group (DMRG) methods via symmetry conserving purification schemes to predict thermal properties of the system with high accuracy.

\textit{Model.---} The proposed mixD $t-J$ model~\cite{Grusdt_tJ} describes mobile fermions whose motion is restricted to be along one dimension, while their spin is coupled through 2D $\mathrm{SU}(2)$ invariant superexchange interactions. The Hamiltonian reads
\begin{equation}
\begin{aligned}
    \hat{\mathcal{H}} =-t \sum_{\sigma, \braket{\mathbf{i}, \mathbf{j}}_x} \hat{\mathcal{P}}_{GW} &\big(\hat{c}_{\mathbf{i}, \sigma}^{\dagger} \hat{c}_{\mathbf{j}, \sigma} + \text{h.c.} \big)\hat{\mathcal{P}}_{GW} \, + \\ & + J \sum_{\braket{\mathbf{i}, \mathbf{j}}} \left( \hat{\mathbf{S}}_{\mathbf{i}} \cdot \hat{\mathbf{S}}_{\mathbf{j}} - \frac{\hat{n}_{\mathbf{i}}\hat{n}_{\mathbf{j}}}{4} \right),
\end{aligned}
\label{eq:mixD_tJ_H}
\end{equation}
where $\hat{c}_{\mathbf{i}, \sigma}^{(\dagger)}$, $\hat{n}_{\mathbf{i}}$ and $\hat{\mathbf{S}}_{\mathbf{i}}$ are fermionic annihilation (creation), particle density, and spin operators on site $\mathbf{i}$, respectively; $\braket{\mathbf{i}, \mathbf{j}}_{(x)}$ denotes nearest neighbor (NN) sites on the 2D square lattice (with subscript $x$ indicating NN sites along the $x$-direction only), and $\hat{\mathcal{P}}_{GW}$ is the Gutzwiller operator projecting out states with double occupancy. 
The mixD $t-J$ model Hamiltonian, Eq.~\eqref{eq:mixD_tJ_H}, features a global $\mathrm{SU(2)}$ spin- and individual $\mathrm{U(1)}$ particle conservation symmetries in each chain $\ell = 1 \dots L_y$. 

The mixD setup in Eq.~\eqref{eq:mixD_tJ_H} can be realized by simulating the Fermi-Hubbard model in the large $U/t$ limit with a strong on-site linear potential along $y$, i.e., $V_{\text{tilt}}(y) = \Delta y$~\cite{Duan2003, Trotzky, Dimitrova}.
The potential gradient $\Delta$ effectively suppresses resonant tunneling along $y$, whereas virtual particle exchanges (and hence spin superexchange) remain intact -- hence realizing the mixD $t-J$ setting, see~\cite{supp_mat} for a more detailed discussion.

\textit{Ground state properties.---} Using DMRG~\cite{Schollwoeck_DMRG, SchollwoeckDMRG2, WhiteDMRG, hubig:_syten_toolk, hubig17:_symmet_protec_tensor_networ}, we calculate ground state properties of the doped mixD $t-J$ model Eq.~\eqref{eq:mixD_tJ_H} on ladder geometries by explicitly implementing the $\mathrm{U(1)}$ particle conservation symmetry in each leg separately. %, combined with a $\mathrm{U(1)}$ symmetry employed in the spin sector. 
The explicit use of tensors with the Hamiltonian's enhanced symmetry greatly decreases computational costs, whereby speedup factors of $\gtrsim 10$ for bond dimensions $\chi \sim 10,000$ are reached. 

\begin{figure}
\centering
\includegraphics[width=1\columnwidth]{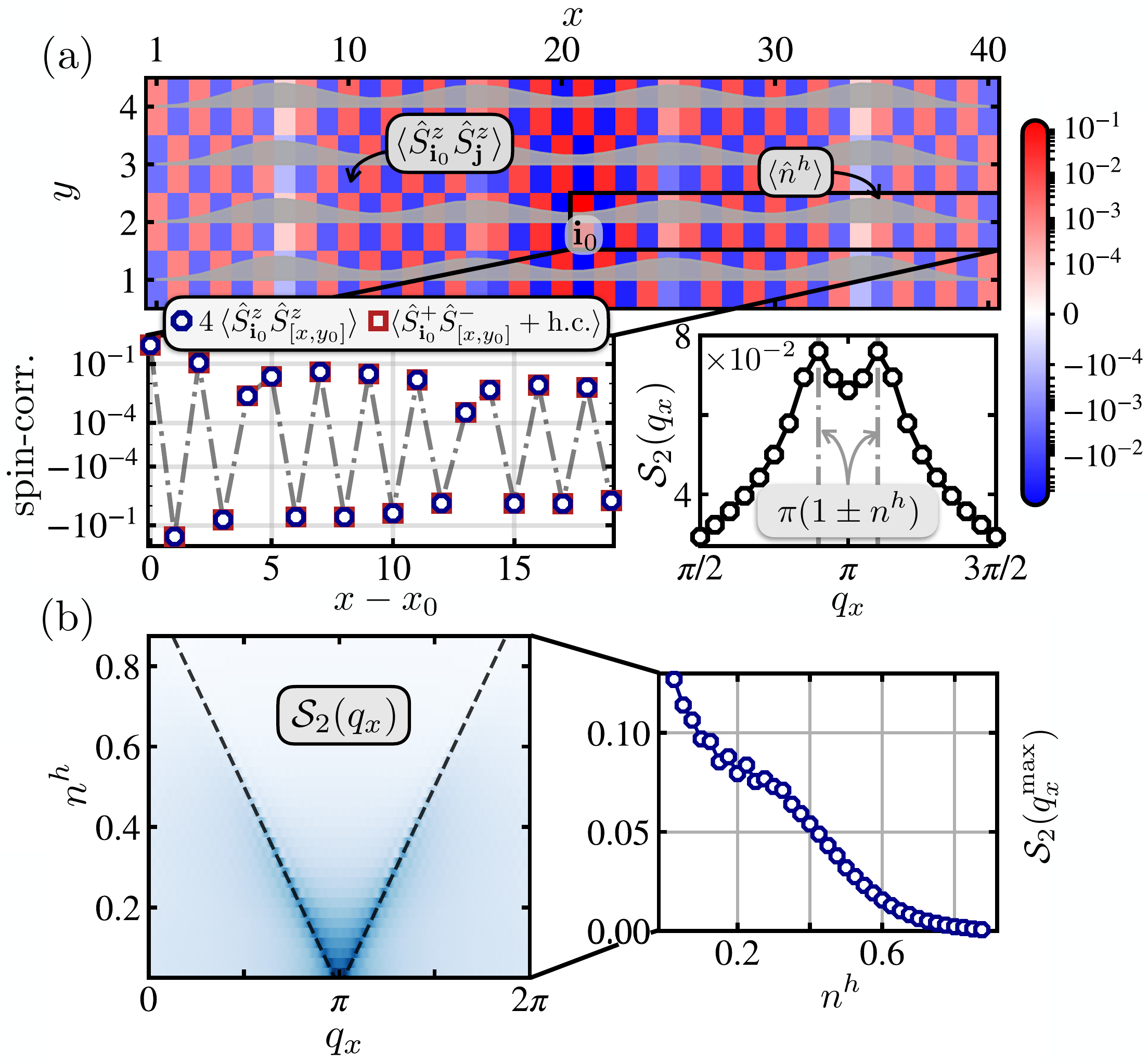}
\caption{Ground state properties. (a) Spin-spin correlations $\braket{\hat{S}^z_{\mathbf{i}_0} \hat{S}^z_{\mathbf{j}}}$ with reference site $\mathbf{i}_0 = [x_0=21, y_0=2]$ for a $40 \times 4$ system with $n^h=0.1$. Boundaries are open (closed) in $x-$ ($y-$) direction. Correlations are color coded using a symmetric logarithmic scale, with linear scaling between $-10^{-4}\dots 10^{-4}$. Average hole densities $\braket{\hat{n}^h_{\mathbf{i}}}$ are shown in grey. Lower left panel: correlation functions $4 \braket{\hat{S}^z_{\mathbf{i}_0} \hat{S}^z_{[x,y_0]}}$, $\braket{\hat{S}^+_{\mathbf{i}_0} \hat{S}^-_{[x,y_0]} + \text{h.c.}}$, which are indistinguishable on the scale of the plot. Dash-dotted grey lines connecting the data points underline the incommensurate peak structure of spin-correlations. Lower right panel: static spin structure factor along $y=2$, Eq.~\eqref{eq:pssf}, with peaks located at $q_x = \pi (1\pm n^h)$.
(b) Left panel: Spin structure factor of the central leg as a function of doping $n^h$ for a $40 \times 3$ system. A narrower system size is chosen to keep numerical costs reasonable. Open boundaries are taken to avoid magnetic frustration of the ladder. Right panel: Peak height $\mathcal{S}_2(q_x^{\text{max}})$ at $q_x^{\text{max}} = (1-n^h)\pi$ as a function of doping $n^h$.}
\label{fig:stripes_gs}
\end{figure}
From now on, we use an equal number of holes in each leg, i.e., $N_{\ell} = N^h$ for all $\ell = 1\dots L_y$, and choose $t/J = 3$. The color coded background in Fig.~\ref{fig:stripes_gs}~(a) shows spin-spin correlations  $\braket{\hat{S}^z_{\mathbf{i}_0} \hat{S}^z_{\mathbf{j}}}$ with fixed reference site $\mathbf{i}_0$ in the center of the second leg for a system of size $L_x \times L_y = 40 \times 4$ with open (periodic) boundaries along $x$ ($y$). Grey filled lines depict local hole densities $\braket{\hat{n}^h_{\mathbf{i}}}$ in each leg. In the ground state, we see clear indications for the formation of fully filled stripes, by observing (i) a periodic modulation of hole densities, and (ii) the appearance of AFM domain walls at positions of maximum hole density. 

The latter is further underlined in the lower left panel of Fig.~\ref{fig:stripes_gs}~(a), where the spin-spin correlations are shown for the central $y=2$ region. Correlations are observed to be incommensurate with the lattice, i.e., the total number of peaks in the spin-correlation function in each ladder leg -- given by $N_p = 18$ in Fig.~\ref{fig:stripes_gs}~(a) -- is incommensurate with the length of the system, $L_x = 40$. This corresponds to a modulation of spin-correlation with wavelength $\lambda = (1-n^h)^{-1}$, where $n^h = N^h/L_x$.
%
%
%This figure is just placed here in the tex file because a figure* is always put to the next page
\begin{figure*}
\centering
\includegraphics[width=0.96\textwidth]{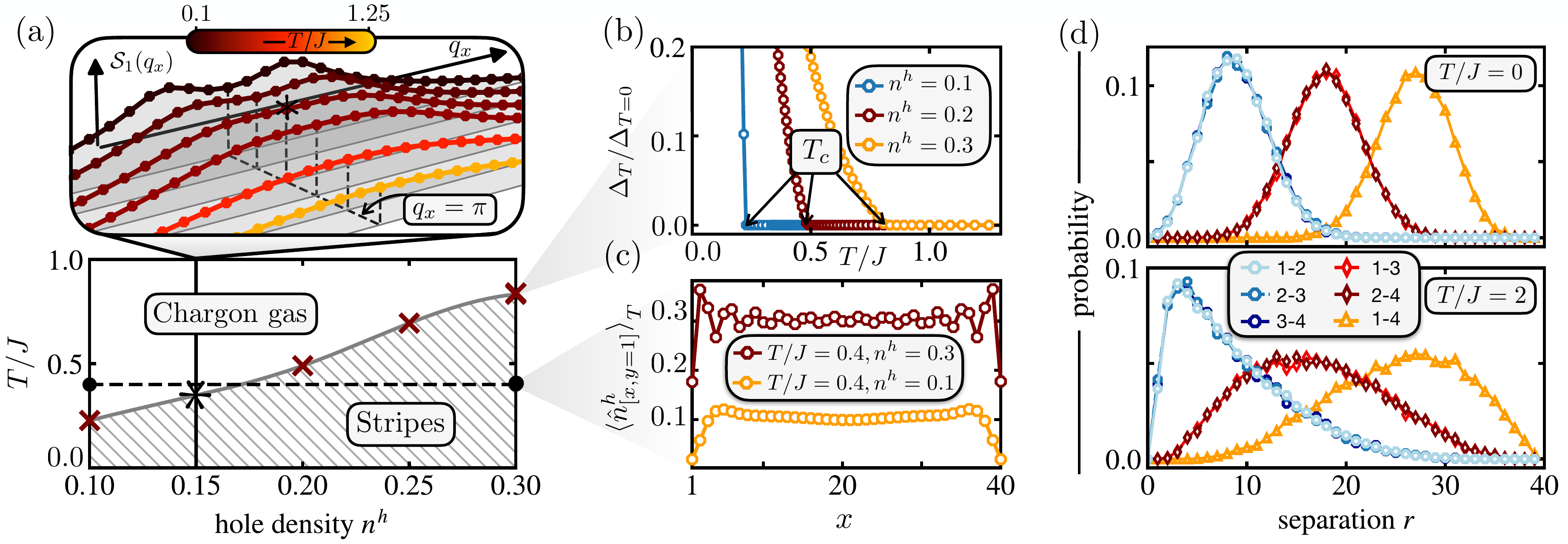}
\caption{Finite temperature properties. (a) Lower panel: Phase diagram of the mixD $t-J$ model for a $40\times 2$ system (OBC). In the stripe phase, charge- and spin-density waves are present, whereas in the chargon gas phase holes can move freely and AFM spin correlations are short-range. We define the critical temperature $T_c(n^h)$ as the point where the double peak of the spin structure factor washes out into a single broad peak, illustrated for $n^h=0.15$ in the upper panel of (a) and marked by an asterisk. Temperatures $T/J = 1.25, 0.83, 0.4, 0.35, 0.25, 0.1$ are shown, ranging from  yellow to dark red upon lowering the temperature. (b) Peak split $\Delta_T/\Delta_{T=0}$, with $\Delta_T = S(q_x^{\text{max}}) - S(\pi)$, as a function of temperature. (c) Hole density profiles $\braket{\hat{n}^h_{[x,y=1]}}_T$ for $n^h = 0.1, 0.3$ at constant temperature $T/J = 0.4$ (solid dots connected by dashed line in (a)). In the stripe phase, clear charge oscillation signals in the hole density profile are visible, whereas in the chargon gas the profile is flat. (d) Full counting statistics of hole distances along a single ladder leg for $n^h = 0.1$ and $T/J=0,2$ (notation $i-j$ corresponds to the distance $r$ between hole $i$ and hole $j$ along $x$). In the stripe phase (upper panel), the probability distributions are symmetric, whereas in the chargon gas (lower panel), hole-hole distance probability distributions acquire long tails. We use 20,000 snapshots of the (thermal) MPS.}
\label{fig:fT}
\end{figure*}
Emerging incommensurate antiferromagnetic order is further revealed in the static spin structure factor along leg $y$,
\begin{equation}
    \mathcal{S}_y(q_x) = \frac{1}{L_x} \sum_{x_1,x_2} \braket{\hat{S}^z_{[x_1,y]} \hat{S}^z_{[x_2,y]}} \text{exp}\big[i q_x (x_1 - x_2)\big],
    \label{eq:pssf}
\end{equation}
which features a double-peak structure at points $q_x^{\text{max}} = \pi (1\pm n^h)$, depicted in the lower right panel of Fig.~\ref{fig:stripes_gs}~(a).
When increasing the doping level, incommensurate magnetic order and stripes persist, however with overall decreasing magnetic order due to the enhanced disturbance by the holes, cf. Fig.~\ref{fig:stripes_gs}~(b). Beyond $n^h \gtrsim 0.5$, no clear signs of stripe formation are visible anymore.

We note that charge-density wave-like correlations are also expected in purely 1D systems with open boundaries. The oscillation amplitudes of such Friedel oscillations away from the edges decay as $r^{-K}$, with $K$ the Luttinger exponent~\cite{Schulz1990}. In contrast, in the mixD setting the amplitude of the charge modulations quickly converges towards a constant plateau, i.e., they are present even deep in the bulk. We explicitly compare the density oscillations in long 1D and mixD systems in~\cite{supp_mat}.

On cylinders of width $L_y = 4$, we evaluate the intra-leg binding energy of two holes $E_b = [E(2) - E(0)] - 2 [E(1) - E(0)]$, where $E(N^h)$ is the ground state energy of a mixD cylinder with $N^h$ holes doped into a single ladder leg. We find almost vanishing binding energies of order $E_b/t \sim \mathcal{O}(10^{-3})$ in our finite-size simulations, strongly supporting the absence of hole-pairing in the mixD $t-J$ model. This is further underlined by features of connected hole-hole correlators, revealing how at short distances, holes strongly repel each other, whereas binding in stripes is favored~\cite{supp_mat}.

\textit{Finite temperature.---} In order to estimate critical temperatures for stripe formation, we use mixed state purification and imaginary time evolution schemes while conserving the system's symmetries~\cite{Nocera2016, Paeckel_time, Feiguin2010, supp_mat}. 
%In particular, we expand the ladder system by introducing auxiliary sites, which act as a finite temperature bath via their entanglement to the physical system. 
%In order to calculate thermal matrix product states, we first generate the infinite temperature, maximally entangled state $\ket{\Psi(\beta=0)}$ in a given symmetry sector~\cite{supp_mat}.
%A pure state in the enlarged system at finite temperature is then calculated by evolving $\ket{\Psi(\beta=0)}$ in imaginary time under the physical Hamiltonian, $\ket{\Psi(\tau)} = e^{-\tau \hat{\mathcal{H}}} \ket{\Psi(\beta=0)}$, where $\tau = \beta/2$ with $\beta$ the inverse temperature. The corresponding mixed state of the physical system is computed by tracing out all auxiliary degrees of freedom when computing expectation values in the physical subset.
In particular, during the time evolution we conserve the particle number in each physical leg $N_{\ell}, \ell=1..L_y$, the total particle number in the auxiliary system $N_{\text{aux.}}^{\text{tot}}$, as well as the total spin $S^{z,\text{tot}}_{\text{phys.+aux.}}$ (the latter allowing for finite total magnetizations of the physical system at finite temperate). This results in a total of $L_y + 2$ symmetries employed by the finite temperature implementation.

%The maximally entangled state needed as a starting point of the imaginary time evolution is generated using the concept of entangler Hamiltonians~\cite{Feiguin2010, Nocera2016}, which we specifically tailor for our ``leg-canonical'' ensemble. Since the maximally entangled state is usually of low bond dimension, we first employ global MPS time evolution schemes to evolve the system away from infinite temperature. Once bond dimensions are sufficiently high, we switch to local approximation methods~\cite{Paeckel_time, supp_mat}. 
Utilizing the enhanced symmetry, we are able to accurately evolve the system down to low temperatures, allowing us to evaluate convergence towards the ground state~\cite{supp_mat}. 
Due to the effective doubling of the width of the system after purification, we restrict the following discussion to systems with $L_y = 2$. Numerical results for wider systems are presented in~\cite{supp_mat}. For the time evolution schemes, we again use maximal bond dimensions $\chi \sim 10,000$.

Results for a $40\times 2$ physical system with open boundary conditions (OBC) are shown in Fig.~\ref{fig:fT}. Starting from high temperatures, we measure the static spin structure factor, localize its peak position $\pm q_x^{\text{max}}$ and calculate the peak split parameter defined by $\Delta_T = \mathcal{S}_y(q_x^{\text{max}}) - \mathcal{S}_y(\pi)$.
The upper panel of Fig.~\ref{fig:fT}~(a) shows the structure factor for various temperatures. At high temperatures, correlations are short-range and in particular commensurate with the lattice, i.e., the structure factor is characterized by a broad peak around $q_x = \pi$ and $\Delta_{T>T_c}$ is strictly zero. Upon lowering the temperature to the critical value $T_c$, a finite split in the structure factor is observed, i.e., incommensurate magnetic features emerge. The transition point is marked by an asterisk in the upper panel of Fig.~\ref{fig:fT}~(a).

Fig.~\ref{fig:fT}~(b) underlines the definition of the critical temperature, where the peak split becomes finite, i.e., $\Delta_{T<T_c}>0$. The corresponding critical temperatures as a function of hole doping are plotted in the lower panel of Fig.~\ref{fig:fT}~(a), for hole densities ranging from $n^h = 0.1...0.3$. Note how critical temperatures are of the order of magnitude $\sim J/2$, rendering the stripe phase significantly more robust against thermal fluctuations in the mixD setting compared to its analog in 2D~\cite{Wietek_stripes, Zheng2017}.

We illustrate the emergence of stripes further by showing the average hole density profile for two different doping levels while keeping the temperature constant, Fig.~\ref{fig:fT}~(c). For $T/J = 0.4$ and $n^h = 0.1$, the hole density forms a flat plateau, i.e. there is no charge order and holes are in a deconfined chargon gas phase (i.e.~ holes are not confined within stripes and move freely through the magnetic background)~\cite{Grusdt_tJz}. In contrast, clear charge oscillations are visible for $n^h = 0.3$, underlining how in the stripe phase both charge and spin density waves are present. By computing the charge structure factor, we have checked that charge order is present over the whole range of temperatures $T<T_c(n^h)$, in fact setting in at slightly higher temperatures than incommensurate magnetic order. This is characteristic for a crossover driven by the charges~\cite{emery1999stripe, Zachar1998}, as is the case for the chargon gas to stripe transition observed here.     

Note that the N\'eel temperature of the $\mathrm{SU(2)}$ symmetric 2D Heisenberg model is strictly zero, however with a magnetic correlation length scaling as $\sim e^{T_0/T}$. Akin to the cold atom antiferromagnet realized in~\cite{Mazurenko_AFM}, we argue that stripe features - that is, the emergence of charge- and spin-density wave patterns - become visible on the length scale of the system size for temperatures below $T_c$. For sufficiently strong magnetic correlations, we expect that true long-range charge order (breaking the discrete translational symmetry of the system) is stabilized at finite temperatures also in the thermodynamic limit $L_x, L_y \rightarrow \infty$.

\textit{Snapshots and hidden correlations.---} In quantum gas microscopy experiments, projective measurements are taken in the Fock basis of the many-body state. These snapshots contain a plethora of information about the system beyond averages and local observables, allowing for further insights into the quantum many-body wave function~\cite{Bohrdt_ML}. Using (thermal) matrix product states, we sample independent snapshots via the perfect sampling approach~\cite{Ferris2012, Buser2022}.

Fig.~\ref{fig:fT}~(d) illustrates the stripe-chargon gas crossover by full counting statistics of hole-hole distances within a single ladder leg. In the stripe phase, probability distributions for hole-hole distances are symmetrically peaked around a maximal distance probability. On the other hand, in the chargon gas phase the discrete probability distributions develop long tails, i.e., mean and maximum are far separated from another. These kind of rare event distributions often govern the physics of the system~\cite{Shu_rare}, here indicating a phase of freely moving, deconfined holes through the magnetic background.

\begin{figure}
\centering
\includegraphics[width=0.9\columnwidth]{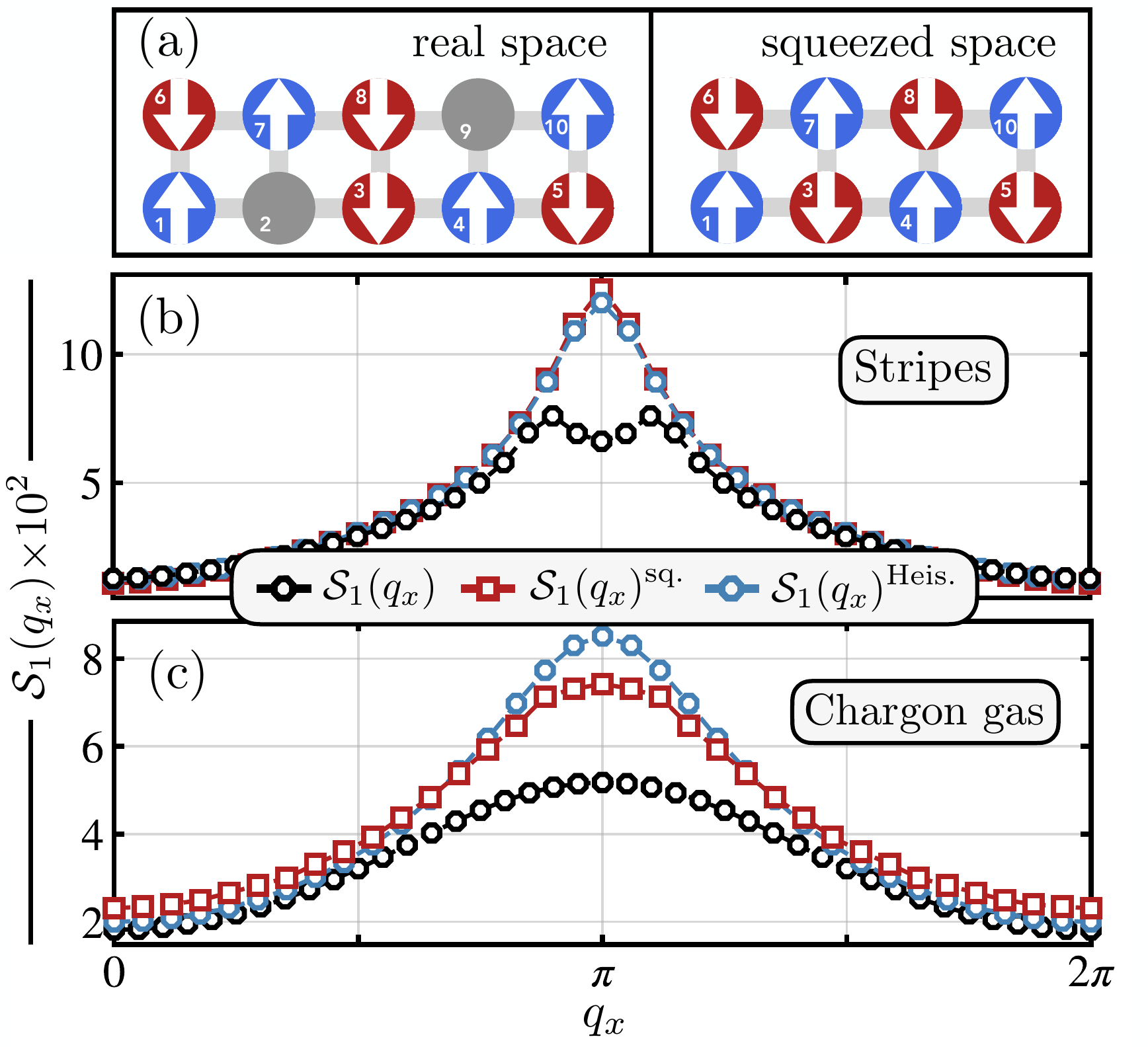}
\caption{(a) Illustration of squeezed space. Left panel: Snapshot of holes moving through a N\'eel background. Right panel: Upon squeezing out the holes, N\'eel order is restored in the magnetic background. (b) Spin structure factor $\mathcal{S}_1(q_x)$ for a mixD $t-J$ system in the striped ground state with $n^h = 0.1$, featuring a double peak structure in real space (black circles). Squeezing snapshots reveals hidden AFM correlations (red squares), being in quantitative agreement with the $J=1$ Heisenberg ground state (blue circles). (c) The same for the chargon gas phase at $T/J = 5/7 \approx 0.71$, $n^h = 0.15$. Upon squeezing, the peak around $q_x=\pi$ becomes considerably sharper. Compared to the pure Heisenberg model with $J=1$ and $T/J = 5/7 \approx 0.71$ (light blue circles), the spin structure in the chargon gas phase has a shifted weight towards FM correlations. For both cases above, the mixD system is of size $40\times 2$.}
\label{fig:HO}
\end{figure}

A further major advantage of the restricted hole motion is that we can uniquely define squeezed space~\cite{Hilker2017, Kruis2004}, where holes are removed (i.e., ``squeezed out'') from each snapshot before measuring observables, cf. Fig.~\ref{fig:HO}~(a).
Spatially separating occupied and unoccupied sites by squeezing allows to analyze the interplay between hole motion and magnetism in more detail, explicitly utilizing non-local information contained in snapshots of the many-body wave function.

Fig.~\ref{fig:HO}~(b) shows the spin structure factor along $y=1$ of a $40\times 2$ mixD $t-J$ model in the ground state (i.e., in the stripe phase) in real (black circles) and squeezed (red squares) space after removing holes from the snapshots.
When transforming ground state snapshots to squeezed space, hidden AFM correlations are revealed, i.e., the double peak structure turns into a sharp peak around $q_x=\pi$. Indeed, when comparing to the pure Heisenberg model~\footnote{We compare the magnetic mixD system in squeezed space with the Heisenberg model, given by the Hamiltonian 
\begin{equation*}
    \hat{\mathcal{H}} = J \sum_{\braket{\mathbf{i}, \mathbf{j}}} \hat{\mathbf{S}}_{\mathbf{i}} \cdot \hat{\mathbf{S}}_{\mathbf{j}},
\end{equation*}
where $J$ is the magnetic coupling constant and $\braket{\mathbf{i}, \mathbf{j}}$ denotes nearest-neighbour pairs on a square lattice.} with $J=1$, both structure factors agree on a quantitative level, showing how the holes confined within the stripes leave the underlying magnetic state in squeezed space almost undisturbed.

In the chargon gas phase, on the other hand, the movement of the holes distorts the magnetic background in a more notable manner. Black circles in Fig.~\ref{fig:HO}~(c) show $\mathcal{S}_1(q_x)$ evaluated via the thermal MPS at $T/J = 5/7 \approx 0.71$. When transforming snapshots to squeezed space, the initially broad peak around $q_x=\pi$ again becomes significantly sharper, showing how AFM correlations are reduced due to the holes' motion through the Mott insulator.
Compared to the Heisenberg model with $J=1$ at $T/J= 5/7 \approx 0.71$ [blue circles in Fig.~\ref{fig:HO}~(c)], the squeezed mixD system has shifted weight from AFM ($q_x=\pi$) to more FM ($q_x=0,2\pi$) correlations. Enhanced FM signals in squeezed space emerge due to the frustrating effect of hole motion on spins in squeezed space, which is analyzed and quantified in detail by some of us in~\cite{HMIF}.

\textit{Discussion.---} In this letter, we presented numerical DMRG results that demonstrate the formation of stable stripes in the $t-J$ model of mixed dimensionality. Above critical temperatures $T_c$ on the order of $J$, we find commensurate, short-range antiferromagnetic correlations together with deconfined holes. Below the critical temperature, incommensurate order as well as charge density waves emerge on long length scales of the numerically accessible system size, i.e., stripes form in the system. Our work extends the strong pairing mechanism proposed in bilayer Hubbard models~\cite{bohrdt2021strong} and realized in mixD ultracold atom setups~\cite{Hirthe2022}, to stabilize stripes at high temperatures. 

Observations of strong AFM correlations in three dimensional (3D) realizations of the FH model~\cite{Hart2015} motivate the existence of resilient sheets of stripes in a possible generalization of the mixD $t-J$ model to 3D, where - in contrast to 2D - true long-range magnetic order appears also at finite temperature.

Recently, the mixD $t-J_z$ model including solely Ising-type interactions has been analyzed~\cite{Grusdt_tJz}. There, an exact mapping revealed an emergent $\mathbb{Z}_2$ lattice gauge structure, which allowed to draw analogies with gauge theories and distinguish phases by emergent properties. The mixD $t-J_z$ model has been shown to exhibit a rich phase diagram when restricted to a single gauge sector (i.e. a fixed AFM N\'eel background), including stripes, a deconfined chargon gas as well as a meson gas, where in the latter holes form pairs at low hole concentrations and temperatures slightly above $T_c$.  
It remains to be analyzed whether a confined phase of mesonic character exists also in the mixD $t-J$ model (including spin fluctuations as well as gauge mixing in comparison to~\cite{Grusdt_tJz}), and if any conclusive connections can be drawn in the context of $\mathbb{Z}_2$ lattice gauge theories.

\textit{Acknowledgments.---} We are thankful for valuable discussions with S. Paeckel, Z. Zhu, S. Mardazad, L. Rammelm\"uller, J. Dicke, F. Palm, M. Kebric, S. Haas, I. Bloch, T. Hilker, S. Hirthe, D. Bourgund, and P. Bojovic. This research was funded by the Deutsche Forschungsgemeinschaft (DFG, German Research Foundation) under Germany’s Excellence Strategy—EXC-2111—390814868, by the European Research Council (ERC) under the European Union’s Horizon 2020 research and innovation programme (grant agreement number 948141), by the FP7/ERC Consolidator Grant QSIMCORR, No. 771891, and by the NSF through a grant for the Institute for Theoretical Atomic, Molecular, and Optical Physics at Harvard University and the Smithsonian Astrophysical Observatory.

\widetext
\appendix
\section{\underline{Supplementary Materials}}

\subsection{Stripes vs. Friedel oscillations}
\begin{figure}[!h]
\centering
\includegraphics[width=0.5\columnwidth]{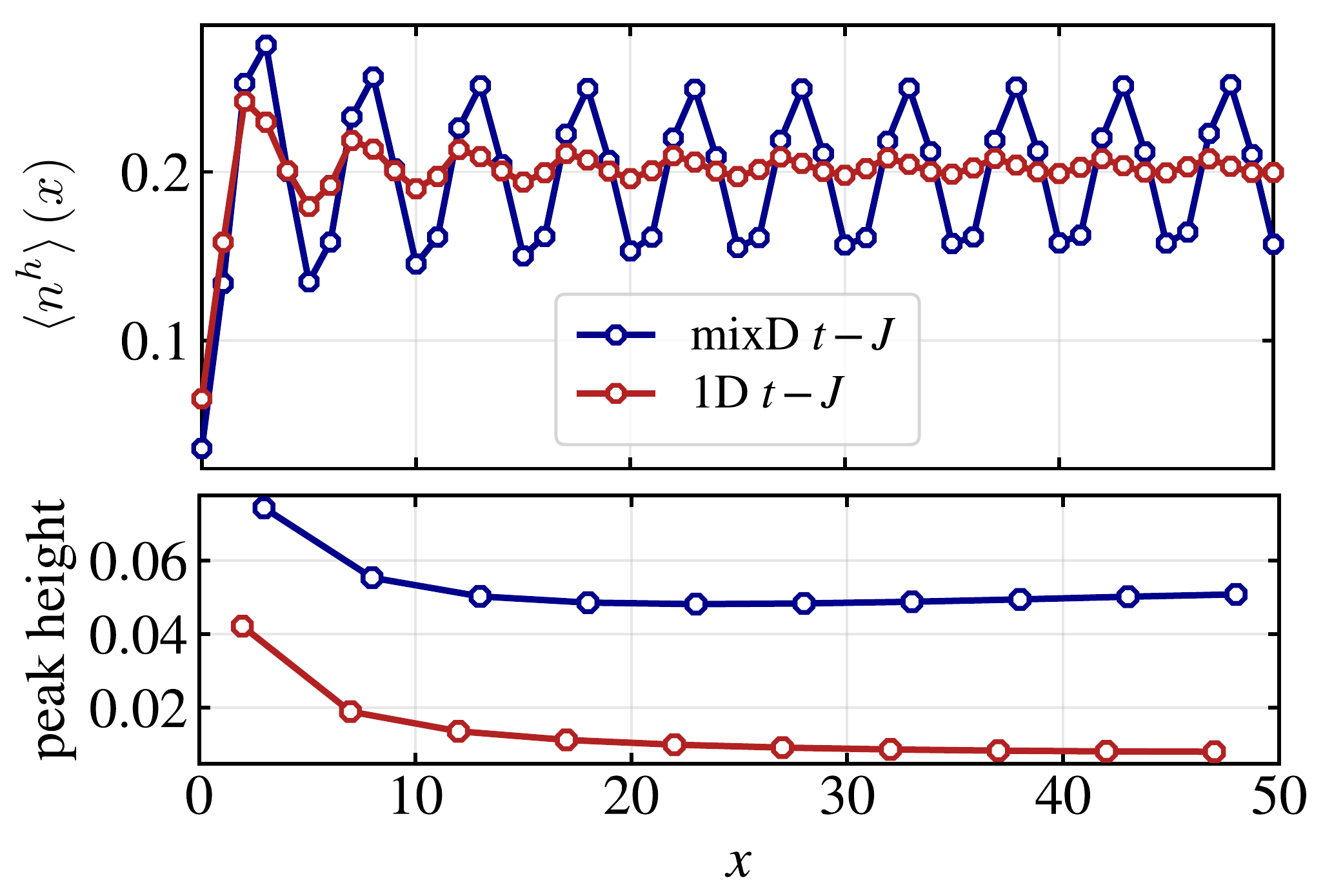}
\caption{Charge density oscillations in 1D and mixD systems. The hole density distribution as a function of position $x$ for both a purely 1D $t-J$ system of length $100$ (red connected dots) as well as a $100\times 3$ mixD $t-J$ system (blue connected dots). In the mixD system, the mean hole density $\braket{\hat{n}^h_{[x,y=2]}}$ of the central leg is shown. Lower panel: charge density oscillation amplitude as a function of peak position $x$. In the 1D case, the oscillations decay with increasing distance from the boundary, whereas constant amplitudes signal stable charge density order in the mixD setting.}
\label{fig:Friedel}
\end{figure}

In the main text, it was argued that the observed charge density wave order in the stripe phase is fundamentally different from 1D Friedel oscillations, the latter emerging due to boundary effects. Note that in the 1D $t-J$ model, charge alternations correspond to $2k_F$ oscillations of free chargons -- which is why we use the term ``Friedel oscillations'' also in the context of the mixD system. In order to clarify the difference between 1D and mixD, consider Fig.~\ref{fig:Friedel}, where the hole density in the ground state is shown for (i) a $100 \times 1$ $t-J$ model and (ii) a $100 \times 3$ mixD $t-J$ system, each with $n^h = 0.2$, i.e., $20$ holes (per leg). 

In the purely 1D system, Friedel oscillations are expected with density modulations decaying like $x^{-K}$ far away from the boundary, with $K$ the Luttinger parameter of the system. Focusing first on the red curve in the upper panel of Fig.~\ref{fig:Friedel}, we observe this typical decay of charge density wave amplitude with increasing distance from the boundary. This is underlined by the lower panel of Fig.~\ref{fig:Friedel}, where we explicitly show the peak height as a function of its position. 

The decay of charge density wave order in 1D is in stark contrast to the oscillations observed in the stripe phase in the mixD system, where the oscillations are seen to quickly converge towards a finite plateau, see the blue curve in Fig.~\ref{fig:Friedel}. This underlines that, though pinning the charge density order, the boundaries are not responsible for the formation of charge density waves. Instead, the magnetic background and arising linear string potential~\cite{Grusdt_tJz} ultimately lead to the formation of stripes. 

\subsection{Absence of hole pairing}
In the main text, is was argued that in the mixD setting, intraleg hole pairing is heavily suppressed, while stripe formation is in turn strongly favored. We illustrate this by calculating the binding energy of two holes doped into the central leg of  $L_y = 4$ cylinders. If $E(N^h)$ corresponds to the ground state energy of the mixD $t-J$ model with $N^h$ holes in a single ladder leg, the binding energy of a hole pair is given by 
\begin{equation}
    E_b = [E(2) - E(0)] - 2 [E(1) - E(0)].
    \label{eq:Eb}
\end{equation}
Hence, for negative binding energies $E_b<0$, a bound state is formed between the two holes. By computing the ground state energies $E(0), E(1), E(2)$, we evaluate the binding energy for a $20\times 4$ mixD $t-J$ system with PBC (OBC) in y- (x-) direction, which we find to be $E_b/t \sim - \mathcal{O}(10^{-3})$. In the thermodynamic limit, in fact, we expect the binding energy to exactly vanish, ultimately supporting the absence of pairing in the mixD $t-J$ model.

\begin{figure}
\centering
\includegraphics[width=0.45\columnwidth]{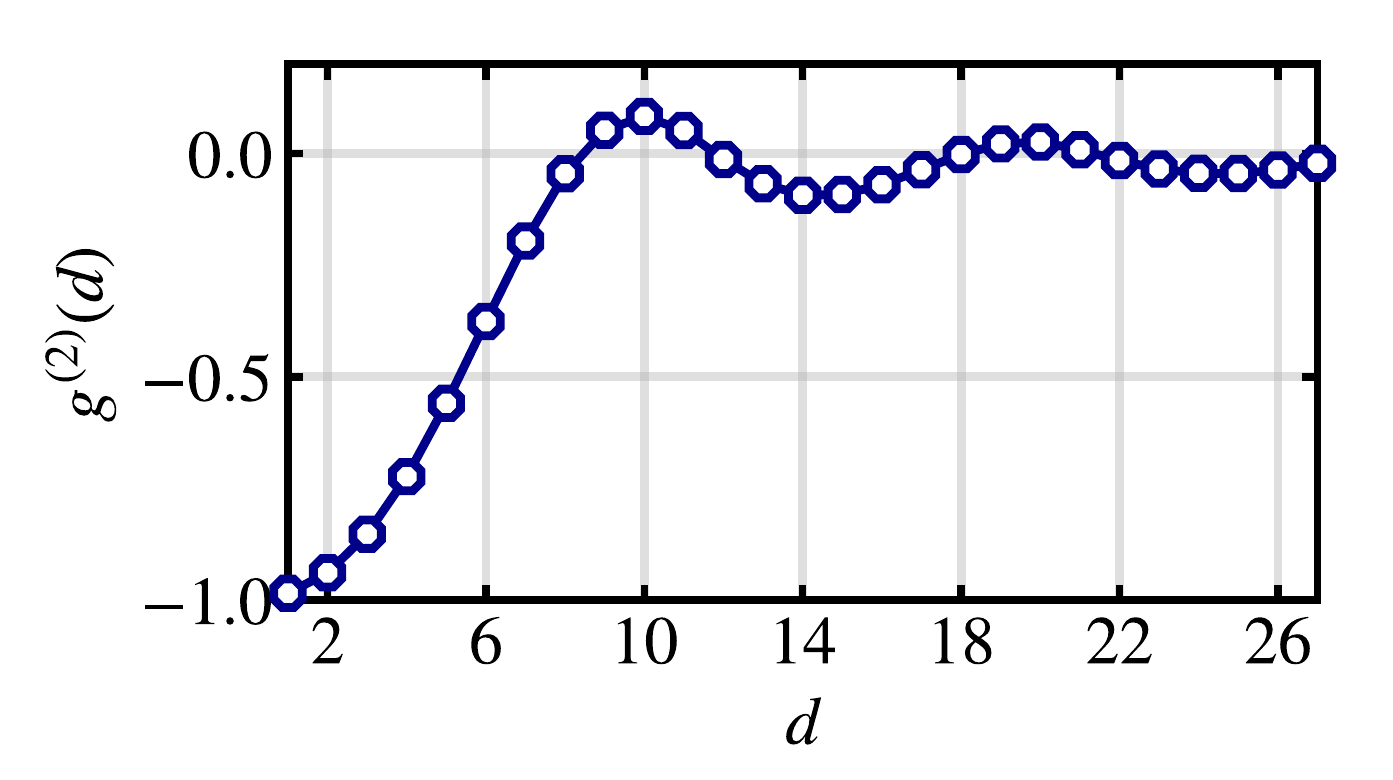}
\caption{The $g_2^{(2)}$ function of holes along a single ladder leg $y=2$, cf. Eq.~\eqref{eq:g2}, for the $40\times 4$ system studied in Fig.~\ref{fig:stripes_gs} in the main text. Values close to $-1$ at short short distances $d \gtrsim 1$ signal the strong tendency of holes to repel each other, and instead arrange themselves in stripes -- signaled by the peak at $d=1/n^h = 10$ and the following modulation of $g_2^{(2)}$. }
\label{fig:Eb}
\end{figure}
To further underline the suppression of hole pairing, we further study the $g_y^{(2)}$ function of the holes along ladder leg $y$, given by
\begin{equation}
    g_y^{(2)}(d) = \left[ \frac{1}{L_x - d} \sum_{i=1}^{L_x -d} \frac{\braket{\hat{n}^h_{i+d} \hat{n}^h_{i}}}{\braket{\hat{n}^h_{i+d}}\braket{\hat{n}^h_{i}}}\right] - 1.
    \label{eq:g2}
\end{equation}
Given a hole at position $[i,y]$, the summand evaluates how likely the existence of a hole at site $[i+d, y]$ is. Thus, if $g_y^{(2)}(d)$ is negative, it is less likely to find holes at positions separated by distance $d$, whereas a positive correlator signals a larger likelihood. Fig.~\ref{fig:Eb} shows $g_2^h(d)$ of the $40 \times 4$ system studied in the main text. Strong negative values close to the lower bound of $g_y^{(2)}(d)$, i.e., $-1$, underlines the tendency of holes avoiding each other. Upon increasing the distance, the connected correlator $g_2^{(2)}(d)$ increases, until reaching a maximum at the mean stripe-stripe distance $d=1/n^h=10$. Visible modulations for further growing distance $d$ signal the presence of a charge-density wave in the ground state, in accordance to the results presented in Fig.~\ref{fig:stripes_gs} in the main text.

\section{Numerical details}
\underline{\textbf{Ground state calculations.}} Due to the additional constraint of hole motion in purely one dimension, the mixD model features an enhanced $\left( \bigotimes_l \mathrm{U(1)}_{N_{\ell}} \right) \otimes \mathrm{U(1)}_{S_z^{\text{tot}}}$ symmetry, which we explicitly employ in our DMRG calculations- i.e., conserved quantum numbers are $[N_{1}, N_2, \dots, N_{L_y}, S_z^{\text{tot}}]$. %This comes with two main advantages. 
\begin{figure}
\centering
\includegraphics[width=0.45\columnwidth]{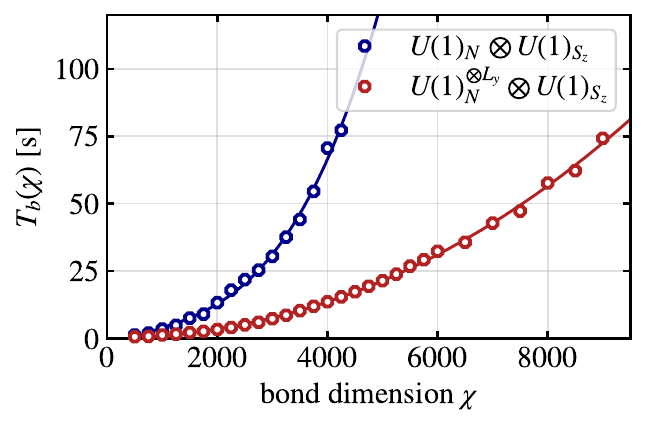}
\caption{CPU time per DMRG step as a function of bond dimension $\chi$ when using a global $\mathrm{U(1)}$ charge conservation symmetry (blue circles) and when including the $\mathrm{U(1)}^{\otimes L_y}$ symmetry in the ground state search (red circles). For $\chi=5,000$, we gain a speedup of a factor $\sim 6$, for $\chi \gtrsim 10,000$, speedup factors are $>10$. Enforcing particle conservation makes large scale computations significantly more efficient, allowing to study the ground state of arbitrary hole configurations with high precision in wide ladders.}
\label{fig:speedup_gs}
\end{figure}

By targeting a smaller subspace in the Hilbert space fulfilling the constraints imposed by the quantum numbers, DMRG sweeps can be done more efficiently during the ground state search. This is exemplified in Fig.~\ref{fig:speedup_gs}, where the average time per sweep as a function of maximum MPS bond dimension $\chi$ is shown with and without using the extended symmetries. We observe a substantial reduction of computational costs -- for instance, the DMRG ground state search is around six times faster for a bond dimension of $\chi=$5,000, and around ten times faster for $\chi=$10,000, the latter being used in the main text for the ground state and finite temperature calculations. In order to guarantee sufficient convergence for ladders of large width, the bond dimension must be increased exponentially in $L_y$. By exploiting the symmetry structure of the mixD setup, we can achieve a high degree of convergence for all correlation functions, see Fig.~\ref{fig:stripes_gs} in the main text. Furthermore, explicit implementation of the separate $\mathrm{U(1)}$ symmetries ensures to always stay in the initially chosen symmetry sector, which allows us to study the mixD system with arbitrary hole configurations with high precision.

For the ground state calculations of the main text, we choose $\chi = 10,000$.
We note that the results are well converged, e.g. the ground state features expectation values $\braket{\hat{S}^z_{\mathbf{i}}} \sim \mathcal{O}(10^{-7})$ and $4\braket{\hat{S}^z_{\mathbf{i}} \hat{S}^z_{\mathbf{j}}} - \braket{\hat{S}^+_{\mathbf{i}} \hat{S}^-_{\mathbf{j}}+ \text{h.c.}} \sim \mathcal{O}(10^{-5})$, which are both expected to vanish for $\mathrm{SU(2)}$ symmetric states. \\ \\

\underline{\textbf{Finite temperature calculations.}} For our finite temperature calculations we use purification schemes, where again all $\mathrm{U(1)}$ symmetries are explicitly conserved. To this end, we enhance the Hilbert space by one auxiliary (often also called ancilla) site per physical site, which allows us to display mixed states in the physical subset of the Hilbert space as pure states on the enlarged space. Thermal expectation values are then computed via
\begin{equation}
    \braket{\hat{O}}_{\beta} = \frac{\braket{\Psi(\beta)|\hat{O}|\Psi(\beta)}}{\braket{\Psi(\beta)|\Psi(\beta)}},
\end{equation}
where $\ket{\Psi(\beta)} = e^{-\beta \hat{H}/2} \ket{\Psi(\beta=0)}$ is the maximally entangled state $\ket{\Psi(\beta=0)}$ evolved in imaginary time $\tau = \beta/2$, and $\mathcal{O}$ is an operator acting on the physical sites only (i.e., all auxiliary degrees of freedom are traced out in the evaluation of the expectation value). Note that this is indeed an exact formulation of the usual form $\braket{\hat{\mathcal{O}}} = \frac{1}{Z} \text{Tr} ( \rho \hat{O} )$, where $Z = \text{dim}(\mathcal{H}) \braket{\psi(\beta)|\psi(\beta)}$ with $\text{dim} (\mathcal{H})$ the dimension of the Hilbert space. Thus, the problem boils down to an imaginary time evolution from the infinite temperature, maximally entangled state $\ket{\Psi(\beta=0)}$. \\

\begin{figure}
\centering
\includegraphics[width=0.5\columnwidth]{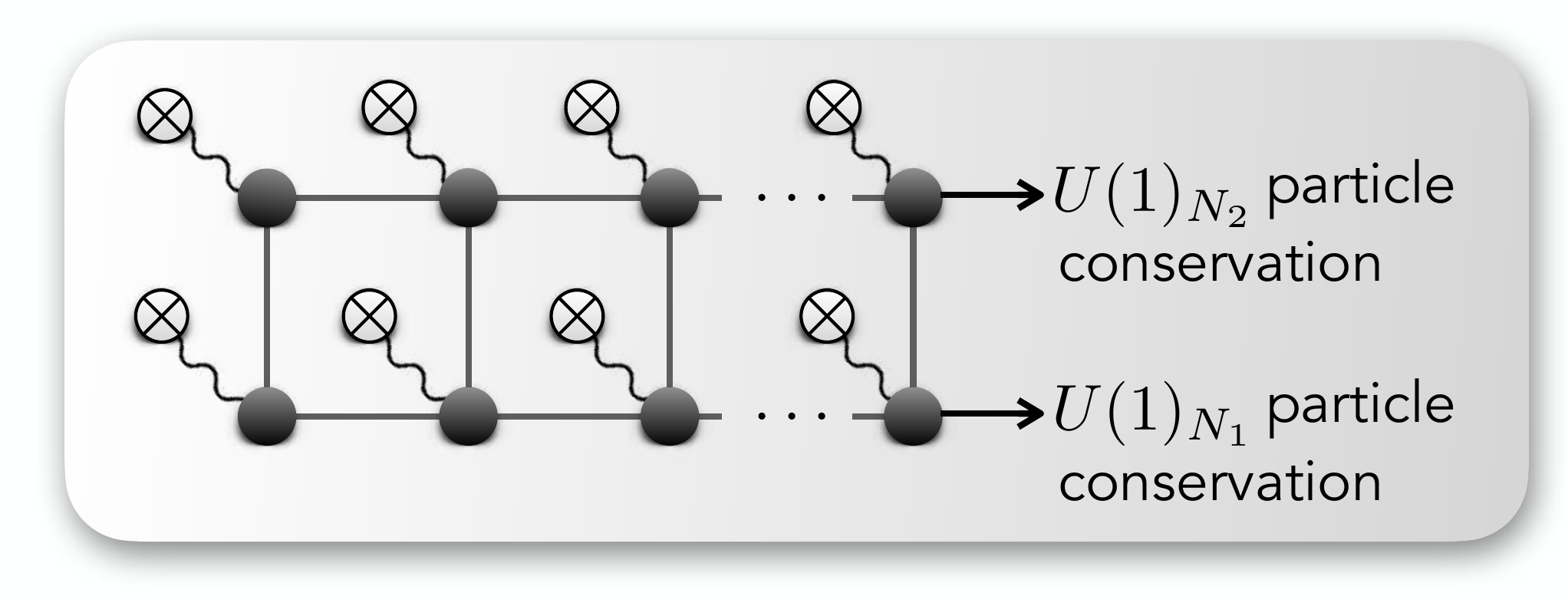}
\caption{Cartoon of the physical and ancilla system to emulate a thermal bath for finite temperature calculations. Physical sites and bonds are illustrated by grey filled circles and black lines, auxiliary sites and their artificial connection to the physical system are shown by crosses and wavy lines. There exists no physical connection between the two Hilbert spaces, nevertheless ancilla sites can act on physical sites implicitly through their entanglement.}
\label{fig:ancilla}
\end{figure}

In order to target our desired subspace also for finite temperature calculations, we need to incorporate the symmetries also in the enlarged Hilbert space, i.e. we would like to perform a calculation in the canonical ensemble while conserving the number of charges in each ladder leg. The structure of the system employed in our calculations is depicted in Fig.~\ref{fig:ancilla}, whereby the particle number in each physical ladder leg is conserved.

Note that we do not conserve the particle number in each ancilla ladder leg (which is,  in fact, redundant and merely increases computational costs when sorting the density matrix). Furthermore, only the total spin of physical and auxiliary system is conserved, i.e. we allow for finite magnetizations in the spin sector at finite temperature. 
In total, conserved quantum numbers are therefore $[N_1, N_2, \dots, N_{L_y}, N_{\text{aux.}}^{\text{tot}}, S_{\text{phys.+aux.}}^{z,\text{tot}}]$, resulting in $L_y+2$ conserved quantities. \\

In the subset of the Hilbert space that fulfils the particle number constraints, the maximally entangled state at infinite temperature reads
\begin{equation}
\label{eq:max_ent}
    \ket{\Psi(\beta=0)} = \prod_{\ell} \hat{\mathcal{P}}_{N_{\ell}} \bigotimes_{i = 0}^{L-1} \left( \ket{0,0} + \sum_{\sigma=\uparrow,\downarrow} \ket{\sigma, \bar{\sigma}} \right).
\end{equation}
Here, $\ell$ is the physical chain index, $L= L_x L_y$ the total number of physical sites in the ladder system, $\{\ket{0}, \ket{\uparrow}, \ket{\downarrow}\}$ is the single particle basis of the $t-J$ model with $\bar{\uparrow} = \downarrow$, $\bar{\downarrow} = \uparrow$, and $\hat{\mathcal{P}}_{N_{\ell}}$ is the projector to the subspace with $N_{\ell}$ dopants in the $\ell^{\text{th}}$ physical chain; the first and second entries in the kets correspond to physical and auxiliary sites, respectively. \\

In order to get the MPS representation of the maximally entangled state, we perform a ground state search of a specifically tailored \textit{entangler Hamiltonian}, given by
\begin{equation}
\label{eq:Hent}
    \hat{\mathcal{H}}^{\text{ent.}} = - \sum_{y} \sum_{\substack{x<x' \\ x\neq x'}} \Delta^{\dagger}_{[x,y]} \Delta_{[x',y]}^{\phantom{0}} + \text{h.c.},
\end{equation}
where $\Delta_{\mathbf{x}}^{\dagger} = \frac{1}{\sqrt{2}} \left( c_{\mathbf{x},\uparrow}^{\dagger} c_{a(\mathbf{x}),\downarrow}^{\dagger} - c_{\mathbf{x},\downarrow}^{\dagger} c_{a(\mathbf{x}),\uparrow}^{\dagger}  \right)$ creates a singlet on site $\mathbf{x}$ paired with its corresponding ancilla site $a(\mathbf{x})$. It is instructive to think of Eq.~\eqref{eq:Hent} as a tight-binding Hamiltonian of hopping singlets on enlarged sites including both the physical and its auxiliary site, from which it can be proven that the ground state of Hamiltonian Eq.~\eqref{eq:Hent} is given by Eq.~\eqref{eq:max_ent} (see e.g.~\cite{Nocera2016} for a discussion of the 1D $t-J$ model).

A technical remark: since the entangler Hamiltonian Eq.~\eqref{eq:Hent} blocks into parts where the physical and ancilla sites are correctly or incorrectly paired in the sense of the maximally entangled ground state, it is important to choose an initial DMRG MPS state that fulfills the pairing for the ground state search of Eq.~\eqref{eq:Hent} (for example, a site in the state $\ket{0, \uparrow}$ is not paired correctly). In the notation of~\cite{Nocera2016}, this means that the initial state should be in $\mathcal{S}_G$. \\

A DMRG run will then yield Eq.~\eqref{eq:max_ent} as the ground state, with eigenvalue $-(L_x -N^h)N^h$. For our system sizes and hole dopings, we notice good DMRG convergence, whereby the variational ground state search results in representations of Eq.~\eqref{eq:max_ent} with typical maximal bond dimensions $\chi_{\text{max}} \sim 100$. We can now employ imaginary time evolution techniques to evolve the state away from $\beta=0$ towards finite temperatures. \\

Since the projected product states are of rather low bond dimension, local approximations of the Hamiltonian (and subsequent exponentiation) will suffer from large projection errors and are of low quality. Hence, we start by employing global methods to cool the system to moderate temperatures, after which the entanglement in the system (and the bond dimension of the thermal MPS) has sufficiently increased to switch to local methods. 

In particular, we start with the global Krylov scheme, where we let the bond dimension expand until $\chi_{\text{max}} = 1024$, after which we switch to the local two-site TDVP method~\cite{Paeckel_time, supp_mat}. For the imaginary time evolution, we choose time steps of $\Delta \tau = 0.05$. We fix weight and truncation cutoffs to $10^{-7}$ and $10^{-8}$, respectively, and use an overall maximum bond dimension of $\chi = 10,000$. When reaching $\tau=2$, we switch to time steps $\Delta \tau = 0.1$. Around the transition, we choose $\Delta \tau = 0.02$. Using the scheme outlined above, we cool the system down to $\tau=4$, which lets us evaluate convergence to the ground state -- a crucial step to assess the quality of the moderate temperature states where the transition from chargon gas to stripes happens. 

\begin{figure}
\centering
\includegraphics[width=\columnwidth]{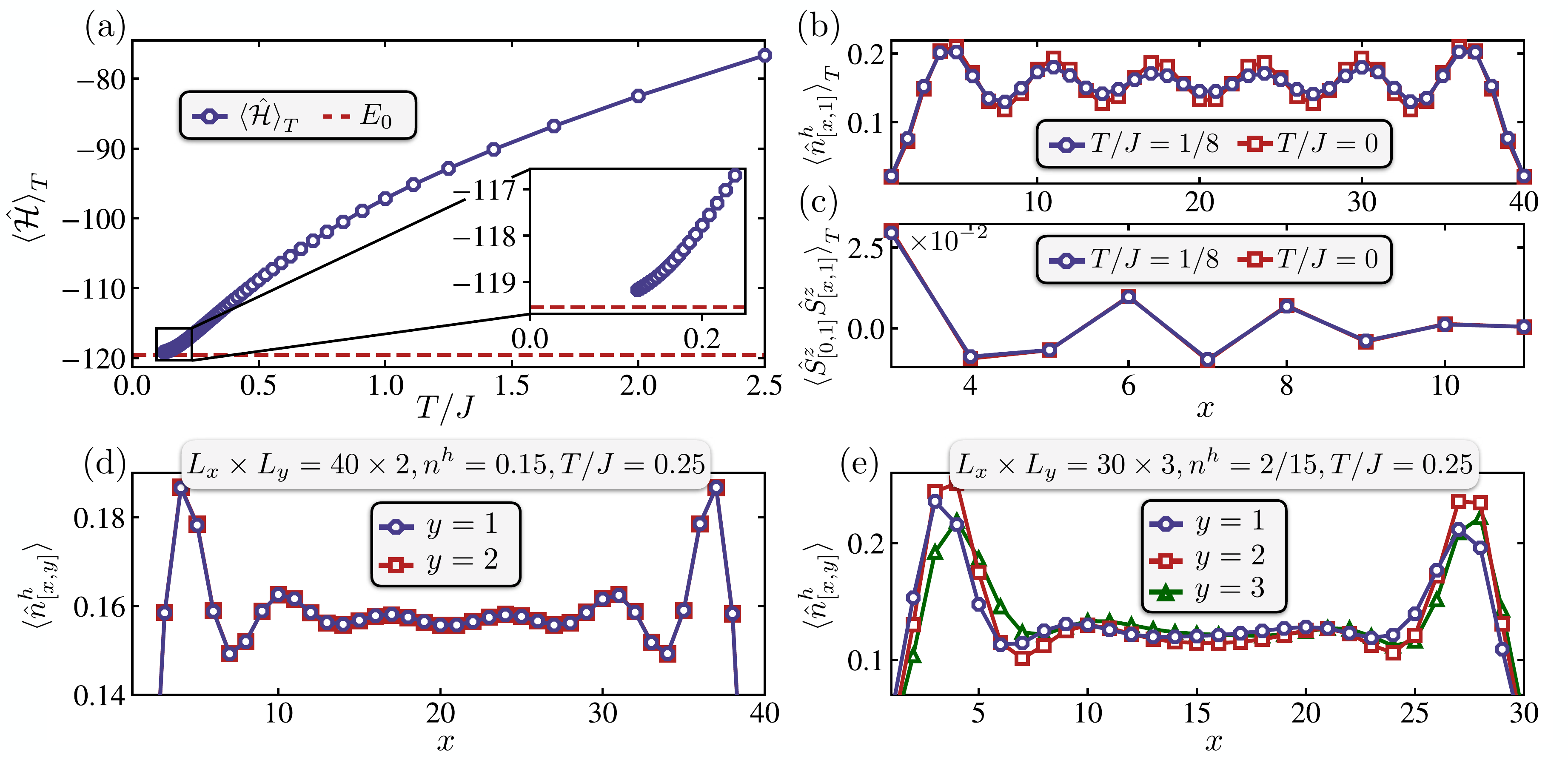}
\caption{(a) Energy $\braket{\hat{\mathcal{H}}}_T$ as a function of temperature $T/J$ for a $40 \times 2$ system at hole doping $n^h = 0.15$. $\braket{\hat{\mathcal{H}}}_T$ approaches the energy $E_0$ from a DMRG ground state search (red line) when $T/J \rightarrow 0$ for a system with identical parameters. For finite temperature calculations, the scheme outlined above is used. (b) Particle density along the lower leg, i.e., $\braket{\hat{n}^h_{[x,1]}}_T$ for temperatures $T/J = 1/8$ (blue dots) and in the ground state $T/J = 0$ (red squares). The density profile converges towards the ground state result for temperatures approaching zero. (c) Spin-spin correlations $\braket{\hat{S}^z_{[1,1]}\hat{S}^z_{[x,1]}}$, again for $T/J = 1/8$ (blue dots) and $T/J = 0$ (red squares). Correlations are seen to converge towards the ground state result. (d)\&(e) Mean hole density profile $\braket{\hat{n}^h_{[x,y]}}_T$ along each ladder leg for a two-leg (d) and three-leg (e) ladder at temperature $T/J=0.25$. Specifically, the system sizes are $L_x \times L_y = 40 \times 2$ for (d) and $L_x \times L_y = 30 \times 3$ for (e), with hole doping $n^h = 0.15$ and $n^h = 2/15$, respectively. In both cases, boundaries are open (OBC). In the two-leg ladder, results along both $y=1$ and $y=2$ are indistinguishable. For the three-leg ladder, charge-density wave patterns are visible, but density distributions are asymmetric. Maximum bond dimensions are $10,000$ and $15,000$ for (d) and (e), respectively. }
\label{fig:convergence_fT}
\end{figure}

Low temperature convergence is illustrated exemplary for a $40\times 2$ system and $N^h = 6$, i.e. $n^h=0.15$, in Fig.~\ref{fig:convergence_fT}. Upon cooling the system from the $\beta=0$ state, the energy $\braket{\hat{\mathcal{H}}}_T$ approaches the energy of a ground state DMRG calculation for $\beta \rightarrow \infty$, see Fig.~\ref{fig:convergence_fT}~(a). In Fig.~\ref{fig:convergence_fT}~(b), the charge density profile $\braket{\hat{n}^h_{[x,1]}}_T$ is shown for the lowest temperature $T/J=1/8$ compared to the ground state results. Good convergence is observed between the $\tau = 4$ and ground state. The same holds for spin-spin correlations, where we show $\braket{\hat{S}^z_{[1,1]}\hat{S}^z_{[x,1]}}$ for $T/J=1/8$ and $T/J = 0$. Again, convergence of ground state and low-temperature purification results is observed. When comparing the density profiles along the two ladder legs, we find that hole densities are indistinguishable for $y=1,2$, cf. Fig.~\ref{fig:convergence_fT}~(d). This underlines good convergence of the two-leg systems.  \\

\begin{figure*}
\centering
\includegraphics[width=0.5\columnwidth]{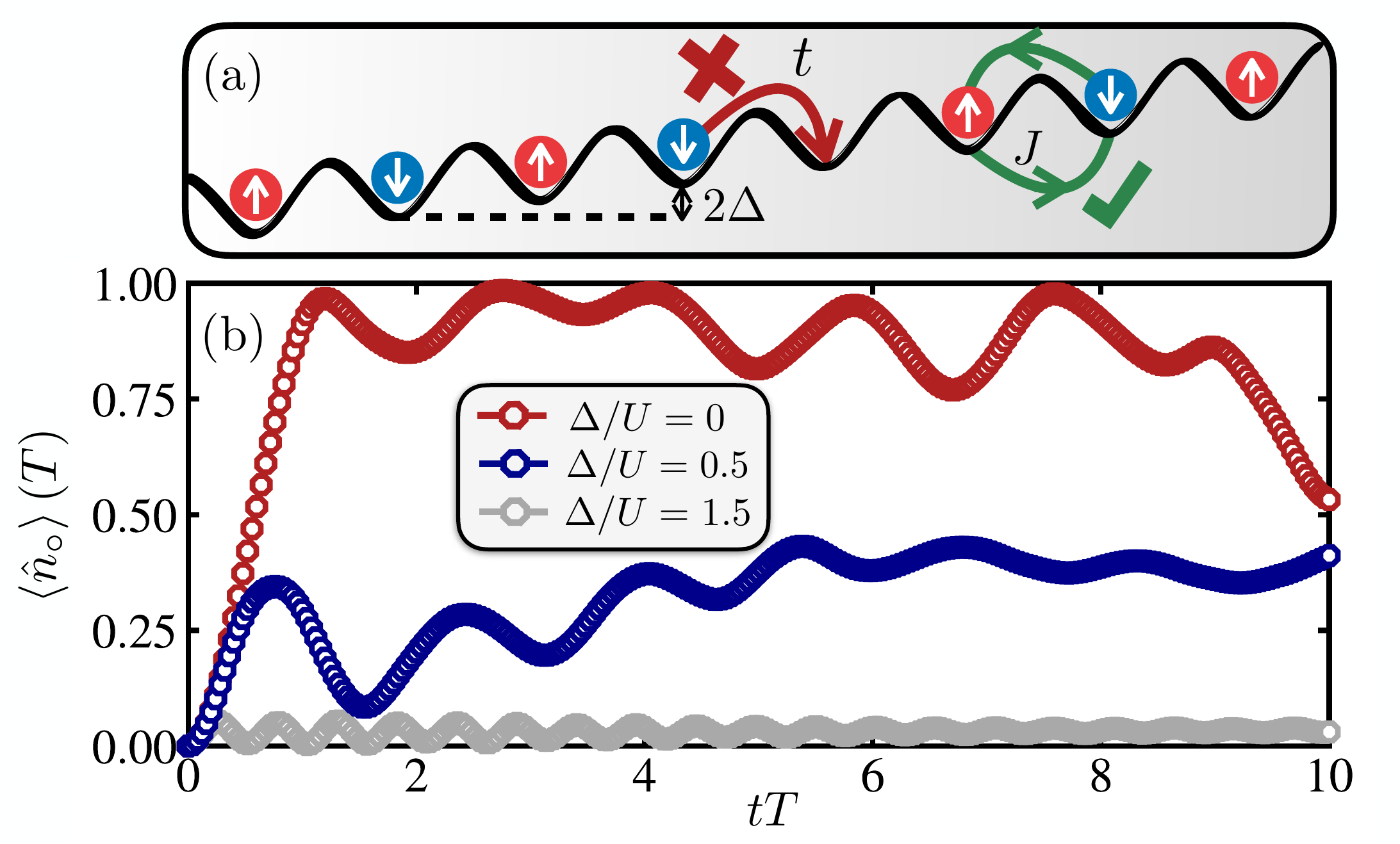}
\caption{(a) Schematic of the space directed tilt procedure (indicated by the slope and gradient background), here shown for a 1D Fermi-Hubbard model in the Mott insulating regime $U/t \gg 1$. While virtual spin exchanges still take place, hopping of spins/holes is suppressed. (b) Time evolution after initializing an eight-site system as shown in (a). Here, $U/t = 8$, and $\Delta/U$ ($\Delta/t$) is varied to values $0,0.5,1.5$ ($0,4,12$). For strong tilts, e.g. $\Delta/U = 1.5$, the density profile is locked in space for exponentially long times, whereas quick delocalization of the hole takes place for both $\Delta/U = 0, 0.5$.}
\label{fig:FH_Delta}
\end{figure*}

Upon cooling the ladder, the entanglement of the system grows rapidly, which results in large bond dimensions of the thermal MPS. With weight and truncation cutoffs as introduced above, we reach the maximal value of  $\chi=10,000$ at around $T/J = 1$. \\
From these magnitude of bond dimensions, it is evident that using the $\mathrm{U(1)}$ symmetries in each ladder leg renders an accurate evaluation of the transition discussed in the main text significantly more feasible. 
Furthermore, we note that purification approaches for systems with $L_y>2$ with the same accuracy is a challenging task. In these cases, using maximally entangled typical thermal states (METTS) might be more advantageous, as demonstrated e.g. in~\cite{Wietek_stripes}. We note, however, that when aiming to reach temperatures of the order $T/J\sim 1$, the number of needed sampled METTS is expected to drastically increase (i.e., long auto-correlation times are expected), making the intermediate temperature regime particularly challenging. 

We implement a $30\times 3$ lattice with open boundaries and run the cooling process identically to the ladders, with maximal bond dimension $\chi = 15,000$. We then analyze the density distribution in the stripe regime. Fig.~\ref{fig:convergence_fT}~(e) shows finite temperature expectation values $\braket{\hat{n}^h_{[x,y]}}_T$ at hole doping $n^h = 2/15$, i.e., $N^h = 4$, along each ladder leg $y=1,2,3$ at temperature $T/J = 0.25$.
Though showing clear signs of charge-density wave features, the resulting hole distributions are not fully converged -- leading to asymmetric densities as in Fig.~\ref{fig:convergence_fT}~(e). In particular, density profiles for $y=1,3$ are expected to be identical, but show visible deviations. We have checked that the spin structure factor shows a split peak with maxima at $q_x = (1-n^h)\pi$ in the three-leg ladder in the stripe regime, but refrain from showing this here. 

\subsection{Experimental realization}
As introduced in the main text, the mixD setup can be realized by simulating the Fermi-Hubbard model in the large $U/t$ limit with a strong $y$-directed on-site potential tilt $\Delta$, i.e., $V_{\text{tilt}}(y) = \Delta y$~\cite{Trotzky, Dimitrova}. 
The potential gradient $\Delta$ here effectively suppresses resonant tunneling along $y$, which results in freezing the projected density on the $y$-axis. On the other hand, virtual particle exchanges (and hence spin superexchange) remain intact - hence realizing the mixD $t-J$ setting. In particular, if $\tilde{t}_{\parallel}$, $\tilde{t}_{\perp}$ denote parallel (along $x$) and perpendicular (along $y$) hopping parameters in the simulated FH model, effective spin couplings in the limit $\tilde{t}_{\parallel}, \tilde{t}_{\perp} \ll U$ are given by~\cite{Duan2003, Trotzky, Hirthe2022} 
\begin{equation}
    J_{\perp} = \sum_{\pm} \frac{2\tilde{t}_{\perp}^2}{U \pm \Delta}, \quad J_{\parallel} = \frac{4 \tilde{t}_{\parallel}^2}{U}.
\end{equation}
Hence, the ratio $J_{\perp}/J_{\parallel}$ reads
\begin{equation}
    \frac{J_{\perp}}{J_{\parallel}} = \left( \frac{\tilde{t}_{\perp}}{\tilde{t}_{\parallel}} \right)^2 \frac{U^2}{U^2 - \Delta^2}.
\end{equation}
For $\tilde{t}_{\perp}<\Delta<U$, tunneling along the perpendicular direction in the effective $t-J$ model is suppressed, $t_{\perp} = 0$, while parallel hopping remains unchanged, i.e., $t_{\parallel} = t = \tilde{t}_{\parallel}$. By coupling multiple 1D FH chains with parameters $\tilde{t}_{\parallel}, U$ via perpendicular couplings $\tilde{t}_{\perp}$ and introducing an offset $\Delta$ between neighboring legs, the proposed mixD $t-J$ model, Eq.~\eqref{eq:mixD_tJ_H} in the main text, can be realized. For instance, for a fixed potential offset $\Delta$, the ratio $\tilde{t}_{\perp}/\tilde{t}_{\parallel}$ can be adapted to tune the system to homogeneous spin-exchange, $J_{\perp} = J_{\parallel}$. \\

To clarify the above arguments, consider Fig.~\ref{fig:FH_Delta}. We simulate the time dynamics of a 1D Fermi-Hubbard model using exact diagonalization with a single hole and half-filled sites elsewhere, in the strong repulsive regime and with varying potential tilts. In particular, we initialize a system with eight sites in the product state $\ket{\uparrow \downarrow \uparrow \downarrow \circ \uparrow \downarrow \uparrow}$ and focus on the time evolution of the particle occupation $\sum_{\sigma} \braket{\hat{n}^{\sigma}_{\circ}}(t)$ of the initially empty site. Upon increasing the potential gradient $\Delta$, the density profile rapidly freezes to the initial configuration, whereas the hole delocalizes in the $\Delta=0$ limit on very short time scales. This illustrates that by strong tilts, resonant tunneling in $y$-direction can be suppressed on large time scales, stabilizing the mixD setup and making it feasible for experimental investigations. \\ \\

\twocolumngrid
\bibliography{robost_stripes_lit}
\onecolumngrid

\end{document}